\documentclass[%
 reprint,
superscriptaddress,
 amsmath,amssymb,
 aps,
prx,
]{revtex4-1}

\usepackage{graphicx}
\usepackage{dcolumn}
\usepackage{bm}
\usepackage{booktabs}  
\usepackage{threeparttable}
\usepackage{multirow}

\begin{document}

\preprint{APS/123-QED}

\title{Multi-label learning for improving discretely-modulated \\continuous-variable quantum key distribution}

\author{Qin Liao}
\affiliation{College of Computer Science and Electronic Engineering, Hunan University, Changsha 410082, China}

\author{Gang Xiao}
\affiliation{College of Computer Science and Electronic Engineering, Hunan University, Changsha 410082, China}

\author{Hai Zhong}%
\affiliation{School of Computer Science and Engineering, Central South University, Changsha 410083, China}

\author{Ying Guo}%
\email{yingguo@csu.edu.cn}\affiliation{School of Automation, Central South University, Changsha 410083, China}

\date{\today}

\begin{abstract}
Discretely-modulated continuous-variable quantum key distribution (CVQKD) is more suitable for long-distance transmission compared with its Gaussian-modulated CVQKD counterpart. However, its security can only be guaranteed when modulation variance is very small, which limits its further development. To solve this problem, in this work, we propose a novel scheme for discretely-modulated CVQKD using multi-label learning technology, called multi-label learning-based CVQKD (ML-CVQKD). In particular, the proposed scheme divides the whole quantum system into state learning and state prediction. The former is used for training and estimating quantum classifier, and the latter is used for generating final secret key. A quantum multi-label classification (QMLC) algorithm is also designed as an embedded classifier for distinguishing coherent state. Feature extraction for coherent state and related machine learning-based metrics for the quantum classifier are successively suggested. Security analysis shows that QMLC-embedded ML-CVQKD is able to immune intercept-resend attack so that small modulation variance is no longer compulsively required, thereby improving the performance of discretely-modulated CVQKD system.

\begin{description}
\item[PACS numbers]
42.50.St
\end{description}
\end{abstract}

\pacs{Valid PACS appear here}
\maketitle


\section{\label{sec:level1} introduction}

For decades, continuous-variable quantum key distribution (CVQKD) \cite{Bennett1984Quantum,S2019P} has been a hotspot in quantum communication and quantum cryptography. It provides an approach to allows two distant legitimate partners, Alice and Bob, to share a random secure key over insecure quantum and classical channels. 

One of the advantage of CVQKD protocol is that the most state-of-art telecommunication technologies can be compatible to CVQKD protocols, so that one may apply CVQKD system to the practical communication network in use. Moreover, CVQKD protocols have been shown to be secure against arbitrary collective attacks, which are optimal in both the asymptotic limit \cite{PhysRevLett.94.020504,PhysRevLett.94.020505,Leverrier:2009cr,Leverrier:2011cu} and the finite-size regime \cite{PhysRevLett.109.100502,PhysRevLett.110.030502}. Recently, CVQKD is further proved to be secure against collective attacks in composable security framework \cite{Leverrier:2015he}, which is the security analysis by carefully considering every detailed step in CVQKD system.

In general, there are two modulation approaches in CVQKD protocol, i.e.,\ Gaussian-modulated CVQKD protocol \cite{PhysRevLett.95.180503,PhysRevA.89.042335} and discretely-modulated CVQKD protocol \cite{Chen:2016dl,PhysRevA.89.042330}. For the first approach, Alice usually encodes key bits in the quadratures ($\hat{p}$ and $\hat{q}$) of optical field \cite{PhysRevLett.97.190503}, while Bob can restore the secret key bits through high-speed and high-efficiency coherent detection techniques. This strategy usually has a repetition rate higher than that of single-photon detections so that Gaussian-modulated CVQKD could potentially achieve higher secret key rate. However, it seems unfortunately limited to much shorter distance due to the problem of quite low reconciliation efficiency in long-distance transmission. For the second approach,  it generates several nonorthogonal coherent states and exploits the sign of the measured quadrature of each state to encode information rather than using the quadrature $\hat{p}$ or $\hat{q}$ itself. This discrete modulation strategy is more suitable for long-distance transmission since the sign of the measured quadrature is already discrete, thereby validating most excellent error-correcting codes even at low signal-to-noise ratio (SNR). However, very small modulation variance is needed for discretely-modulated CVQKD to keep it safe, which largely restrains its further development \cite{Leverrier:2009cr}. For example, the optimal modulation variances in four-state protocol and eight-state protocol are approximately 0.3 \cite{Zhang:2012jc} and 0.25 \cite{GuoPerformance}, respectively. 

On the other hand, most of the existing CVQKD protocols, both Gaussian-modulated protocols and discretely-modulated protocols, are based on information-theoretical techniques, and comply with the similar pattern in which raw key is firstly generated and followed by the post-processing including reconciliation, parameter estimation, error-correcting and privacy amplification \cite{Grosshans:2002gm}. Although a number of excellent CVQKD protocols have been proposed such as measurement-device-independent (MDI) CVQKD protocols \cite{Liao:2018de,Pirandola2015High,PhysRevA.89.052301,Guo:2017du,PhysRevA.97.052327}, plug-and-play (PP) CVQKD protocols \cite{Choi:2016bj,Huang:2016km}, unidimensional CVQKD protocols \cite{Usenko:2015ds,Wang:2017it,Liao:2018id}, entangled-source-in-middle (ESIM) CVQKD protocols \cite{PhysRevA.87.022308,Guo:2017ce}, phase-encoded and basis-encoded CVQKD protocols \cite{PhysRevA.98.012340,pengpra} etc., one may hardly make a further breakthrough for CVQKD due to the limitation of its traditional pattern and techniques.

In recent years, multi-label learning, which is a kind of machine learning technology \cite{2012ML}, has been widely applied and has shown its powerful impact upon diverse research fields such as artificial intelligence, image processing, data mining and so on. Multi-label learning allows each instance belongs multiple labels simultaneously, therefore it is more in line with the objective description of the real world. Meanwhile, some recent works about improving CVQKD using machine learning-based method \cite{PhysRevApplied.12.014059,Liao2018Long,Liu:2018dk} have shown the feasibility. In this paper, we further propose a novel scheme, called ML-CVQKD, for discretely-modulated CVQKD using multi-label learning technology. This scheme is quite different from traditional discretely-modulated CVQKD, it divides the whole quantum system into state learning and state prediction. The former is used for training and estimating quantum classifier, and the latter is used for generating final secret key. In the meantime, a well-behaved quantum multi-label classification (QMLC) algorithm is designed as an embedded classifier, which is not only quite suitable for the proposed system, but also beneficial for improving CVQKD performance. Feature extraction for coherent state is also presented to better fit the input requirement of QMLC, and machine learning-based metrics for quantum classifier are subsequently suggested. We show that the proposed scheme waives the necessity of small modulation variance in discretely-modulated CVQKD, so that QMLC can exploit larger and reasonable variance to more precisely predict unknown coherent state, thereby enhancing the performance of discretely-modulated CVQKD system. Moreover, the proposed scheme can be applied to the existing system without deploying any extra equipment, and it is more economic than other existing CVQKD protocols in terms of practicality.

This paper is structured as follows. In Sec. \uppercase\expandafter{\romannumeral2}, we demonstrate the proposed scheme for discretely-modulated CVQKD. In Sec. \uppercase\expandafter{\romannumeral3}, we elaborate the principle of embedded QMLC algorithm. Analysis and discussion are presented in Sec. \uppercase\expandafter{\romannumeral4} and final conclusions are drawn in Sec. \uppercase\expandafter{\romannumeral5}.

\section{Multi-label learning-based CVQKD}

In this section, we first briefly retrospect the traditional process of CVQKD system, and then give detailed description of the proposed multi-label learning-based CVQKD scheme.

\subsection{Traditional process of CVQKD}
Since F. Grosshans and P. Grangier put forward the initial Gaussian-modulated coherent state (GG02) protocol \cite{PhysRevLett.88.057902}, scholars have proposed plenty of improved schemes. Subsequently, A. Leverrier et al. further modified the process by taking finite-size effect into account \cite{Leverrier:2010es}, rendering the traditional process become a standard for most CVQKD systems.  

\begin{figure}
\centering
\includegraphics[width=3.4in,height=1.16in]{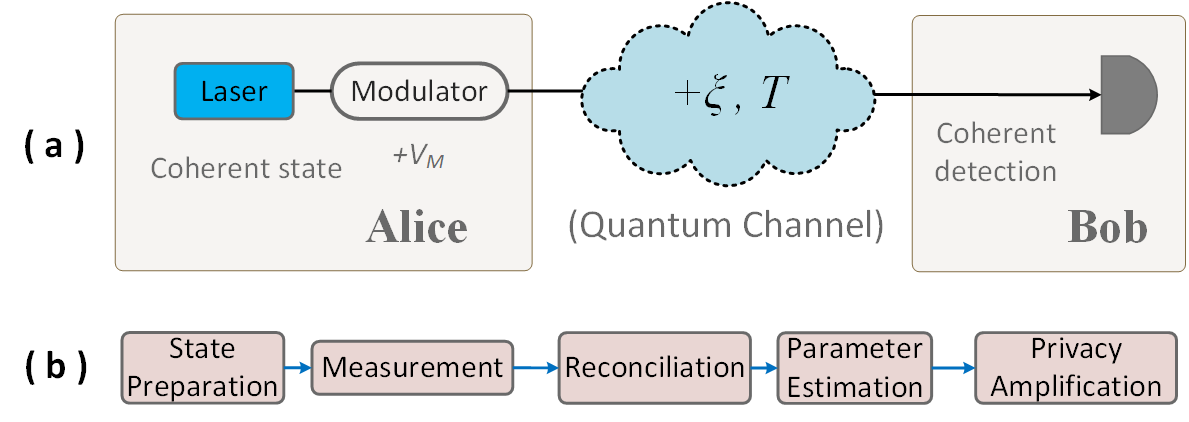}
\caption{(a) Schematic diagram of prepare-and-measurement CVQKD protocol; (b) Traditional process of CVQKD protocol.}
\label{fig:ML}
\end{figure}

Fig. \ref{fig:ML}(a) shows the basic prepare-and-measurement (PM) model for traditional CVQKD, and its process is shown in Fig. \ref{fig:ML}(b) \cite{Leverrier:2015he}, we now explain each step as follows.

\textbf{State Preparation} Alice first prepares a modulated coherent state, and sends it to Bob through an untrusted quantum channel.

\textbf{Measurement} Bob then measures the incoming state with coherent detection so that Alice and Bob share two correlated sets variables, i.e. raw key.

\textbf{Reconciliation} One first needs to discretize the yielded raw key if Gaussian modulation is used, and then a linear error correcting code, usually a low density parity check (LDPC) code \cite{Jiang:2017hu}, is exploited to automatically reconciliate the data between Alice and Bob. 

\textbf{Parameter Estimation} Bob sends a part of bits of information to Alice that allow her to infer the characteristic of the quantum channel and compute the covariance matrix of quantum system. 

\textbf{Privacy Amplification} Alice and Bob apply a random hash function to their respective strings so that they can obtain two identical strings, i.e., secret key.

The above process is based on information theory, so that one needs to correct the erroneous code and evaluate the quality of quantum channel, which makes reconciliation and parameter estimation become the most crucial steps. Therefore, a part of resources such as the generated raw key, storage and computing of devices have to be inevitably sacrificed for these two steps. Moreover, small modulation variance is needed to keep discretely-modulated CVQKD protocol safe, otherwise eavesdropper may perfectly launch the intercept-resend attack without detection \cite{Leverrier:2011cu}.

\subsection{Process of ML-CVQKD}

\begin{figure}
\centering
\includegraphics[width=3.2in,height=3.2in]{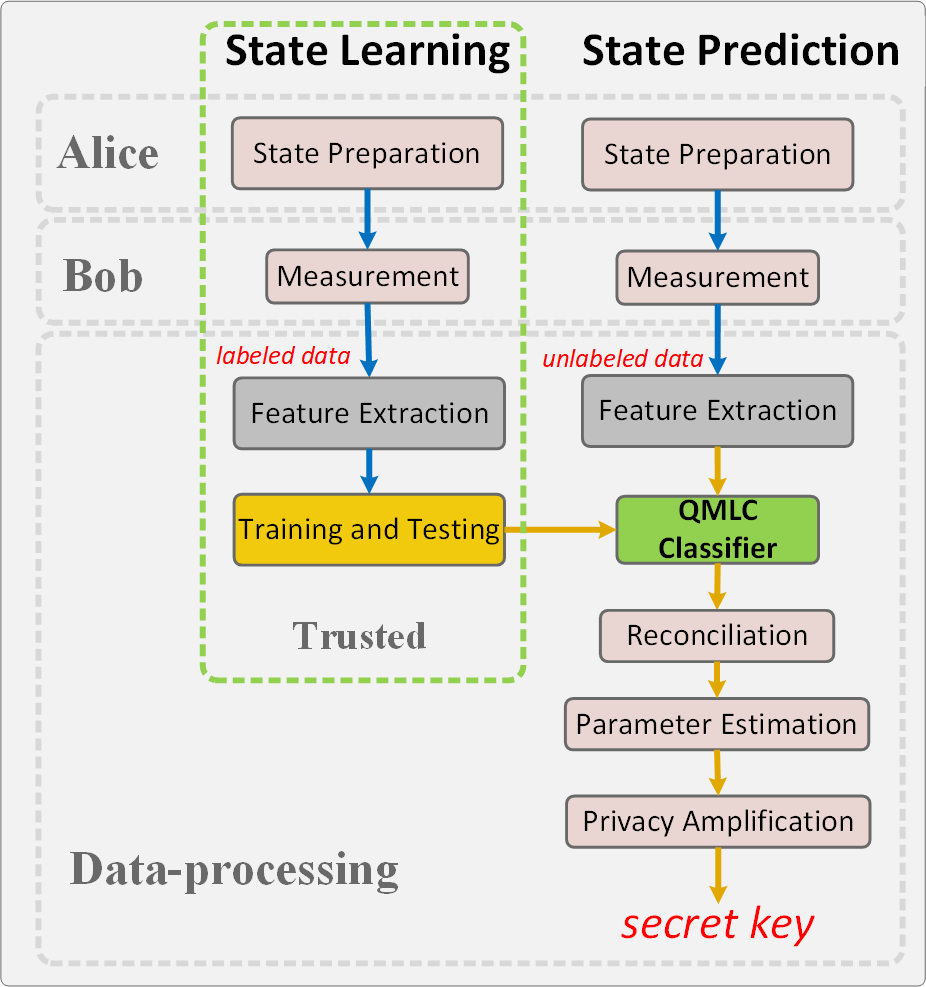}
\caption{Scheme for ML-CVQKD. It is composed of two parts, state learning for modeling and evaluation, and state prediction for classification.}
\label{fig:MLS}
\end{figure}

Multi-label learning involves three modules, i.e., training, testing and prediction. Each module is responsible for respective task, i.e. modeling, evaluation and classification. Specifically, to construct the classifier, a training set is first used for learning the classification rules. Subsequently, another set of data, called testing set, is exploited to evaluate the classifier's performance. Finally, the trained classifier can be used for predicting unknown data if the evaluation is passed. Inspired by the process of multi-label learning, ML-CVQKD is proposed, which includes two parts: state learning and state prediction. As shown in Fig. \ref{fig:MLS}, we explain each step as follows.

\textbf{State Learning} 

\textit{Step 1} Alice first prepares modulated coherent states, and sends them to Bob through noisy and lossy quantum channel.

\textit{Step 2} Bob measures the incoming states with coherent detector thereby obtains the measurement results. Note that these results are similar to, but not identical to the modulated information sent by Alice, this is because the transmitted signals are inevitably distorted by several negative effects such as channel noise and loss.

\textit{Step 3} Bob then extracts features from the obtained labeled coherent states, these features are prepared to training and testing quantum classifier.

\textit{Step 4} After collecting sufficient featured data, Bob divides them into two datasets, i.e. training set and testing set. The former is used for training classifier, and the latter is used to evaluate the classifier's performance. Finally, a well-behaved quantum classifier for ML-CVQKD is prepared if testing is passed. 

It is worthy to noticing that the above steps has to be done without eavesdropping, that is to say, assuming state learning cannot be compromised. This assumption can be implemented through security monitoring of the communication system when initially deploying the proposed scheme. Once the classifier has been trained successfully, one does not need to perform state learning repeatedly. The system is ready for generating secret key.

\textbf{State Prediction} 

\textit{Step 1} Alice prepares modulated coherent states and sends them to Bob through untrusted quantum channel.

\textit{Step 2} Bob obtains the measurement results (unlabeled) by measuring the received states with coherent detector.

\textit{Step 3} Bob first extracts features from the obtained unknown data, these features are subsequently used as input data for the prepared classifier.

\textit{Step 4} Bob classifies the input data using the well-behaved classifier, so that he can predict the state that Alice sent to him. After many rounds of prediction, Alice and Bob share a string of key.

\textit{Step 5} A linear error correcting code, usually a low density parity check (LDPC) code, is applied to automatically correct the data between Alice and Bob.

\textit{Step 6} Bob sends a part of bits of information to Alice that allow her to infer the characteristic of the quantum channel and compute the covariance matrix of quantum system. 

\textit{Step 7} To further enhance the security, Alice and Bob apply a random hash function to their respective strings. Finally, they respectively obtain two identical strings, i.e. secret key.

The steps in state prediction look similar to traditional CVQKD process, however, it is quite different. First of all, in structure, the steps of feature extraction and classifier are added to the process, these two steps are the point of ML-CVQKD scheme. Secondly, the data format is different, coherent state is represented by several robust features rather than quadratures itself, these proper features are conducive to improve the performance of quantum classifier. Moreover, by dividing the whole process into state learning and state prediction, ML-CVQKD is able to immune the intercept-resend attack, so that small modulation variance is no longer required in discretely-modulated CVQKD thereby improving the performance of quantum communication system, the detailed analysis is given at Sec. \uppercase\expandafter{\romannumeral4}.

\section{Quantum multi-label classification}

\begin{figure}
\centering
\includegraphics[width=2.4in,height=1.95in]{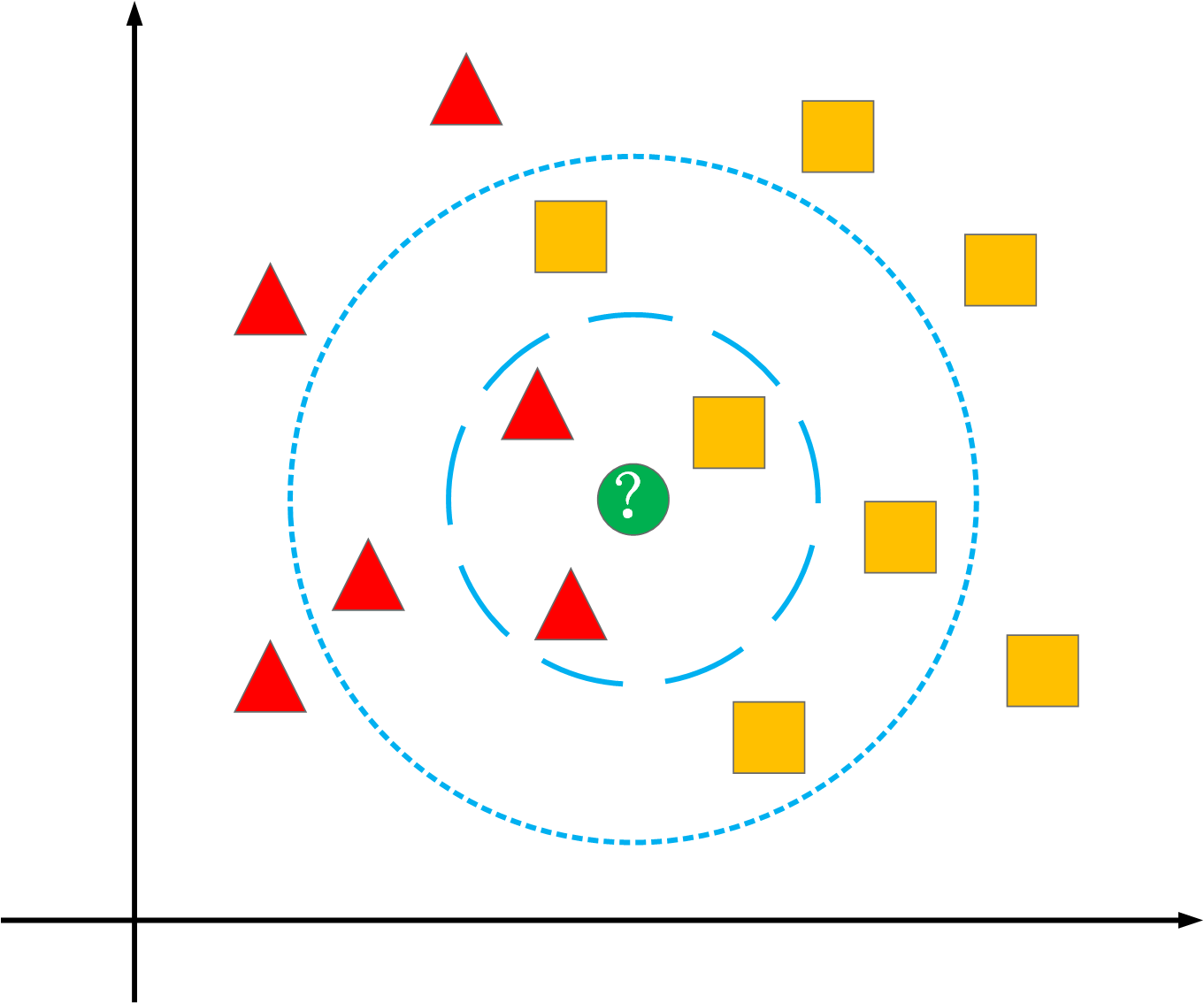}
\caption{The $k$-nearest neighbor classification algorithm in feature space. Green circle denotes an unknown data point, triangles and rectangles belong to red class and yellow class, respectively. Blue dashed circle represents $k=3$ while blue dotted circle represents $k=7$.}
\label{fig:KNN}
\end{figure}

Quantum multi-label classification (QMLC) is derived from a traditional lazy learning approach called $k$-nearest neighbor ($k$NN) \cite{Zhang2007ML}. It selects $k$ nearest neighbors for each unknown data point. Based on the number of neighboring data belonging to each possible class, maximum a posteriori (MAP) principle can be exploited to allocate the label to the unknown data point. Fig. \ref{fig:KNN} depicts an example of $k$NN approach in feature space. The green circle denotes an unknown data point, while triangles and rectangles represent the labeled data points which respectively belong to red class and yellow class. The green circle will be assigned to the red class for $k=3$, since two of the 3-nearest labeled data points belong to the red class while only one point belongs to the yellow class. Similarly, the green circle will be labeled as yellow class for $k=7$.

\begin{figure}
\centering
\includegraphics[width=3.5in,height=2.9in]{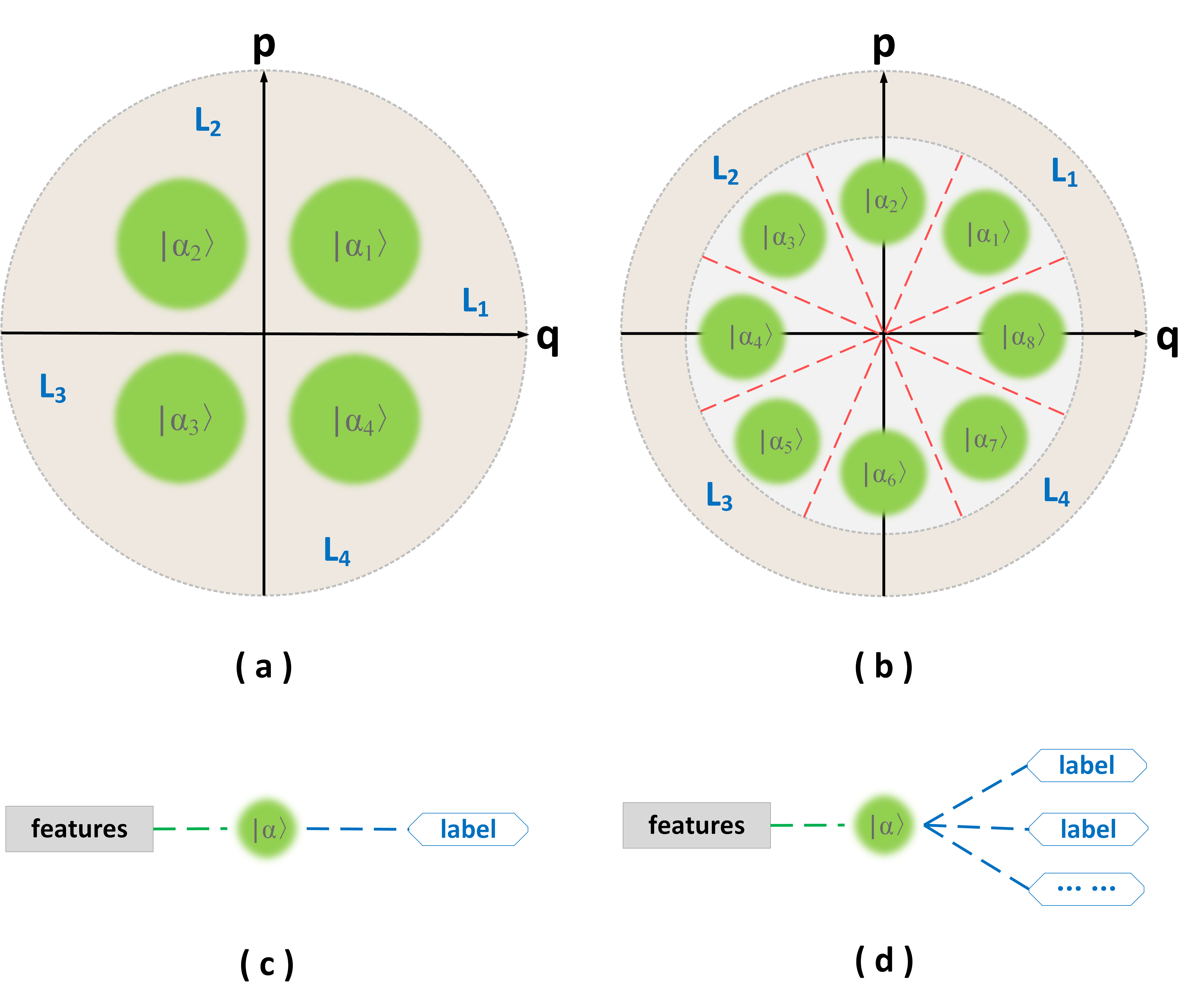}
\caption{(Top) Phase space representation of coherent states with (a) QPSK and (b) 8PSK modulation. Each coherent state is assigned only one label in QPSK or multiple (at least one) labels in 8PSK. (Bottom) Quantum machine learning model for (c) single-label learning and (d) multi-label learning.}
\label{fig:PS}
\end{figure}

$kNN$ classifies unknown data in feature space which enlightens us that coherent states can probably be classified in its phase space. As shown in Fig. \ref{fig:PS}(a), the phase space is divided into several regions, which are labeled as $L_i$ ($i=1,2,3,4$) according to their located quadrant. We find that each QPSK-modulated coherent state belongs to a single label, which can be deemed quantum single-label learning problem depicted in Fig. \ref{fig:PS}(c). However, with the development of modulation technique, single-label learning is not suitable to address high-dimensional modulation problem. As an example, Fig. \ref{fig:PS}(b) shows the phase space representation for coherent states with 8PSK modulation. Thereinto, some of 8PSK-modulated coherent states, such as $|\alpha_2\rangle$, $|\alpha_4\rangle$, $|\alpha_6\rangle$ and $|\alpha_8\rangle$, simultaneously belong to multiple labels, which can be generalized into multi-label learning problem depicted in Fig. \ref{fig:PS}(d). 

In fact, single label is a special case of multiple labels, so that both can be described by the model of multi-label learning. In what follows, we detail the proposed QMLC algorithm for addressing the generalized multi-label CVQKD model. Without loss of generality, we consider the algorithm for the eight-state CVQKD since it is the simplest multi-label modulation scheme. We note that the proposed  QMLC algorithm can also be extended for other complicated modulation schemes.

\subsection{Feature extraction for coherent state}

As known, feature extraction is an important data-preprocessing step in machine learning field, since a set of suitable features would significantly enhance classification performance. The more features are extracted, the more details about the object can be obtained. However, there is few apparent features to describe a modulated coherent state, except for a few attributes such as $p$-quadrature, $q$-quadrature and modulated variance $V_M$.

\begin{figure}
\centering
\includegraphics[width=3.5in,height=1.5in]{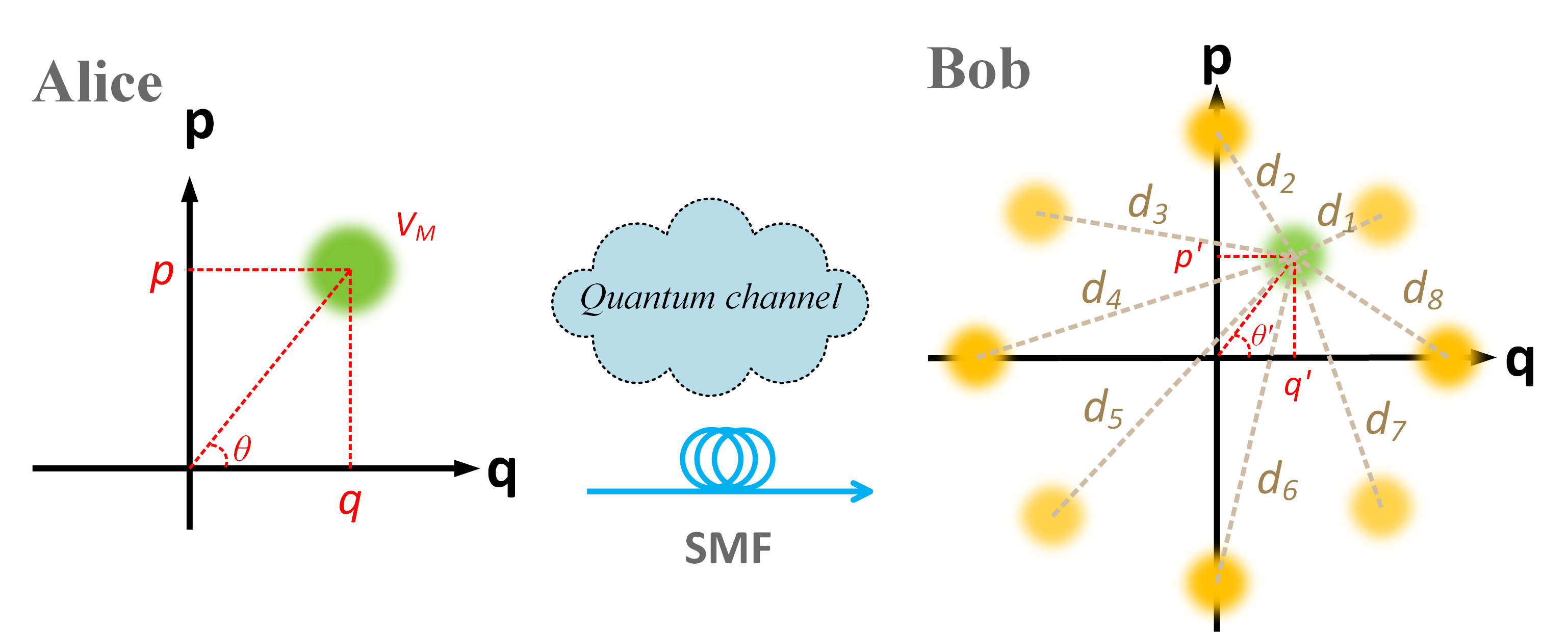}
\caption{Feature extraction for 8PSK-modulated coherent state. (Left) Alice sends the modulated coherent state through the noisy and lossy channel. (Right) Bob receives the transmitted coherent state and extracts its distance features. Green dot denotes signal state, and yellow dot denotes virtual state, i.e. reference state.}
\label{fig:feature}
\end{figure}

To solve the above-mentioned problem, we construct a set of distance features for each coherent state. As shown in Fig. \ref{fig:feature}, Alice sends a modulated coherent state through an untrusted quantum channel (usually a single mode fiber, SMF), and then the transmitted coherent state is received by Bob. Note that the transmitted state is no longer identical with its initial modulated state due to the phase drift ($\theta'\neq\theta$) and energy attenuation ($\sqrt{p'^2+q'^2}<\sqrt{p^2+q^2}$) caused by the imperfect channel noise and loss. Subsequently, a number of virtual states (we named them reference states) are set for calculating the similarities of the transmitted state and reference states. In particular, the similarity can be measured by \textit{Euclidean metric}, which is the straight-line distance between two points in Euclidean space \cite{Deza2009Encyclopedia}. In the Cartesian coordinates, we assume $\textbf{y}=(y_1,y_2,...,y_n)$ and $\textbf{z}=(z_1,z_2,...,z_n)$ are two points in Euclidean $n$-dimensional space, and the distance $d$ between \textbf{y} and \textbf{z} is given by
\begin{equation}
\begin{aligned}
d(\textbf{y},\textbf{z})=\sqrt{\sum_{i=1}^n(y_i-z_i)^2}.
\end{aligned}
\end{equation}
Specifically, in the 2-dimensional phase space we have
\begin{equation}
\begin{aligned}
d_w(\textbf{t},\textbf{r}) & = \sqrt{(p'-p_{r_w})^2+(q'-q_{r_w})^2},
\end{aligned}
\end{equation}
where $w$ is the number of reference state, $\textbf{t}=(p',q')$ and $\textbf{r}=(p_{r},q_{r})$ are the respective Cartesian points of transmitted state and $r$-th reference state. After that, we can extract a set of feature vectors $\bm{d}=(d_1,d_2,...,d_w)$ for better description of the transmitted states.

As mentioned above, reference states are a set of virtual states that do not really exist, and hence one does not need to prepare them at Bob's side. In general, reference states are set to be identical with initial modulated states, which can help us to investigate the influence of imperfect channel on transmitted state.

\subsection{Quantum multi-label classifier}

After extracting robust features, these features are subsequently used as input data of classifier for state learning.  Assuming $\mathcal{X}=\mathbb{R}^d$ is $d$-dimensional data space, and $\mathcal{Y}=\{y_1,y_2,...,y_l\}$ is label space containing $l$ categories. A training set is given by $\mathcal{D}=\{(\bm{x}_i,Y_i)|1\leq i\leq m\}$, where $\bm{x}_i\in\mathcal{X}$ is $d$-dimensional attribute vector $(x_{i1},x_{i2},...,x_{id})^\mathrm{T}$ and $Y_i\subseteq\mathcal{Y}$ is a set of labels to which $\bm{x}_i$ belongs. The task of learning system is to find a multi-label classifier $h(\cdot):\mathcal{X}\to2^{\mathcal{Y}}$. Namely,  for a given threshold function $t:\mathcal{X}\to\mathbb{R}$, it renders $h(\bm{x})=\{y|f(\bm{x},y)>t(\bm{x}),y\in\mathcal{Y}\}$.

Let $|x\rangle$ be an unlabeled coherent state, and $\mathcal{N}(|x\rangle)$ denotes the subset of $k$ nearest coherent states of $|x\rangle$ in training set. The following statistic will be calculated as
\begin{equation}
\begin{aligned}
C_j=\sum_{(|x^*\rangle,Y^*)\in\mathcal{N}(|x\rangle)}[\![y_j\in Y^*]\!],
\end{aligned}
\end{equation}
where $C_j$ counts the number of neighbors of $|x\rangle$ belonging to the $j$-th category $y_j$ $(1\leq j\leq l)$. Assuming $H_j$ represents the event that coherent state $|x\rangle$ has label $y_j$, then $\mathbb{P}(H_j|C_j)$ denotes the posteriori probability, where $H_j$ is true under the condition that $C_j$ is the labeled data in $\mathcal{N}(|x\rangle)$ have label $y_j$. Accordingly, $\mathbb{P}(\bar{H_j}|C_j)$ denotes the posteriori probability which $H_j$ is false under the condition that $C_j$ labeled data in $\mathcal{N}(|x\rangle)$ with label $y_j$. Let $f(|\bm{x}\rangle,y_j)=\mathbb{P}(H_j|C_j)/\mathbb{P}(\bar{H_j}|C_j)$, the quantum multi-label classifier can be expressed by 
\begin{equation}
\begin{aligned}
h(|\bm{x}\rangle)=\{y_j|\mathbb{P}(H_j|C_j)/\mathbb{P}(\bar{H_j}|C_j)>t(|\bm{x}\rangle), 1\leq j\leq l\}.
\end{aligned}
\end{equation}
In other words, unlabeled coherent state $|x\rangle$ can be assigned to category $y_j$ when posteriori probability $\mathbb{P}(H_j|C_j)$ is greater than $t(|x\rangle)\cdot\mathbb{P}(\bar{H_j}|C_j)$.

Specifically, based on \textit{Bayesian theorem} \cite{Liao2018Long}, function $f(|\bm{x}\rangle,y_j)$ can be rewritten as
\begin{equation}
\begin{aligned}
f(|\bm{x}\rangle,y_j)=\frac{\mathbb{P}(H_j|C_j)}{\mathbb{P}(\bar{H_j}|C_j)}=\frac{\mathbb{P}(H_j)\cdot\mathbb{P}(C_j|H_j)}{\mathbb{P}(\bar{H_j})\cdot\mathbb{P}(C_j|\bar{H_j})},
\end{aligned}
\end{equation}
where $\mathbb{P}(H_j)$ and $\mathbb{P}(\bar{H}_j)$ respectively represent the prior probability that event $H_j$ is true or false, $\mathbb{P}(C_j|H_j)$ and $\mathbb{P}(C_j|\bar{H_j})$ respectively represent the conditional probability of $C_j$ labeled coherent states in $\mathcal{N}(|x\rangle)$ with label $y_j$ under the condition that event $H_j$ is true or false. 

The probabilities in Eq. (5) can be estimated by frequency counting in training set. In particular, prior probabilities can be calculated by
\begin{equation}
\begin{aligned}
\mathbb{P}(H_j)&=\frac{s+\sum_{i=1}^m[\![y_j\in Y_i]\!]}{s\times2+m} \quad (1\leq j\leq l),
\end{aligned}
\end{equation}
and
\begin{equation}
\begin{aligned}
\mathbb{P}(\bar{H_j})&=1-\mathbb{P}(H_j) \quad (1\leq j\leq l),
\end{aligned}
\end{equation}
where $s$ is a smoothing parameter controlling the weight of uniform prior distribution during probability estimates, and it usually set to 1 for \textit{Laplace smoothing}.

Different from prior probability, the estimation of conditional probabilities in Eq. (5) is complicated. For the $j$-th category $y_j$ $(1\leq j\leq l)$, we calculate two arrays $\varsigma_j$ and $\bar{\varsigma}_j$, each of which contains $k+1$ elements given by
\begin{equation}
\begin{aligned}
\varsigma_j[r]=\sum_{i=1}^m[\![y_j\in Y_i]\!]\cdot[\![\psi_j(|\bm{x}_i\rangle)=r]\!] \quad (0\leq r\leq k),
\end{aligned}
\end{equation}
and
\begin{equation}
\begin{aligned}
\bar{\varsigma}_j[r]=\sum_{i=1}^m[\![y_j\notin Y_i]\!]\cdot[\![\psi_j(|\bm{x}_i\rangle)=r]\!] \quad (0\leq r\leq k),
\end{aligned}
\end{equation}
where
\begin{equation}
\begin{aligned}
\psi_j(|\bm{x}_i\rangle)=\sum_{(|x^*\rangle,Y^*)\in\mathcal{N}(|x_i\rangle)}[\![y_j\in Y^*]\!].
\end{aligned}
\end{equation}
 $\psi_j(|\bm{x}_i\rangle)$ counts the number of neighbors that belong to category $y_j$ in $k$ nearest neighbors of the $i$-th coherent state. Correspondingly, $\varsigma_j[r]$ counts the number of coherent states that belong to category $y_j$ themselves and exactly have $r$ neighbors which belong to category $y_j$ in $k$ neighbors, while $\bar{\varsigma}_j[r]$ counts the number of coherent states that does not belong to category $y_j$ and exactly have $r$ neighbors which belong to category $y_j$ in $k$ neighbors. Consequently, the conditional probabilities in Eq. (5) can be calculated by
\begin{equation}
\begin{aligned}
\mathbb{P}(C_j|H_j)=\frac{s+\varsigma_j[C_j]}{s\times (k+1)+\sum_{r=0}^{k}\varsigma_j[r]},
\end{aligned}
\end{equation}
and
\begin{equation}
\begin{aligned}
\mathbb{P}(C_j|\bar{H}_j)=\frac{s+\bar{\varsigma}_j[C_j]}{s\times (k+1)+\sum_{r=0}^{k}\bar{\varsigma}_j[r]},
\end{aligned}
\end{equation}
where $1\leq j\leq l$ and $0\leq C_j\leq k$. Finally, a well-behaved quantum multi-label classifier $h(|\bm{x}\rangle)$ is obtained by the successful state learning.

Comparing with state-discrimination detector reported by our previous work \cite{Liao2018Long}, the proposed QMLC algorithm has several advantages. The most obvious merit is that QMLC has the ability to address the model of multi-label learning, thereby it is suitable to the high dimensional modulation strategy. In essence, QMLC belongs a part of data-processing in ML-CVQKD, so that it can be ran without any extra device or component. Moreover, the QMLC-embedded ML-CVQKD can further improve the performance of quantum communication system, we give the detailed analysis in next section.

\section{Analysis and discussion}

In this section, we elaborate the performance and security of the proposed QMLC-embedded ML-CVQKD system. We first interpret the prepared data after feature extraction, and then show the performance analysis of QMLC with several machine learning-based metrics. Security analysis and practicality are subsequently presented.

\subsection{Data preprocessing}
\begin{figure}
\centering
\includegraphics[width=3.5in,height=3.0in]{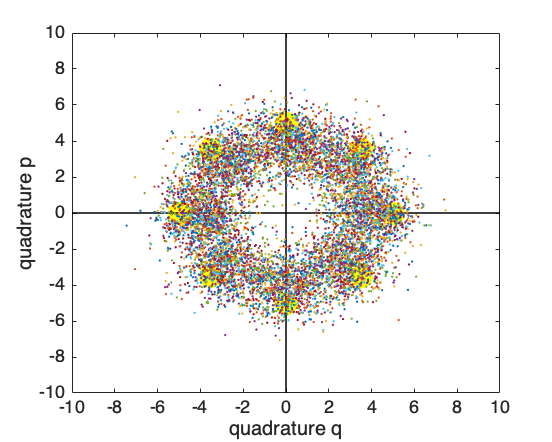}
\caption{$10^{4}$ data points (colored) of 8PSK-modulated coherent state in phase space after passing 20km quantum channel. Yellow dots denote reference states. Modulation variance $V_m=50$ and excess noise $\xi=0.01$.}
\label{fig:sample}
\end{figure}

\begin{figure}
\centering
\includegraphics[width=3.6in,height=2.8in]{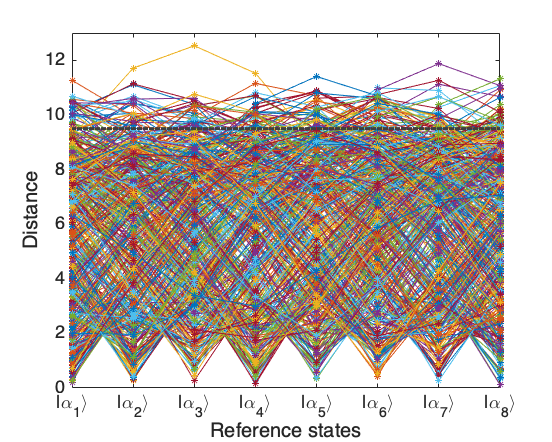}
\caption{Feature vectors of data points after extracting eight-dimensional distance values. Black dashed line represents filter threshold function.}
\label{fig:features}
\end{figure}

As known, quantum channel of  the fiber-based one-way quantum key distribution can be deemed a mapping function which can be described as \cite{Qi2007Experimental}
\begin{equation}
q'=\sqrt{T}(q\cos{\varphi_0}+p\sin{\varphi_0})+\varepsilon,
\end{equation}
\begin{equation}
p'=\sqrt{T}(p\cos{\varphi_0}-q\sin{\varphi_0})+\varepsilon,
\end{equation}
where $\varphi_0=|\theta-\theta'|$ is the phase drift during transmission and $\varepsilon$ is Gaussian $(0,N_0+T\xi)$ distribution. Fig. \ref{fig:sample} shows $10^4$ data points of 8PSK-modulated coherent state in phase space after passing 20km fiber-based quantum channel. Due to the impact of channel loss and noise, the transmitted states are distributed in phase space with a certain probability distribution.
As can be seen, however, these chaotic points with initial format are hardly distinguished so that cannot be directly used as input data for QMLC algorithm. After feature extraction, Fig. \ref{fig:features} shows that these coherent states are mapped into an eight-dimensional vector by calculating the Euclidean distance between each data point and each reference state. We observe that most distance values of feature vectors are located range from 0 to 9.5, while a few feature vectors contains high distance values. These high-value feature vectors are corresponding to the edge outliers in Fig. \ref{fig:sample}, which leads to performance reduction. Therefore, a threshold function can be used to filter the high-value feature vectors, feature vectors whose feature value beyonds black line in Fig. \ref{fig:features} should be discarded for performance improvement. 

\subsection{Performance on machine learning-based metrics}
ML-CVQKD takes advantage of multi-label learning-based technology to predict unknown signal state, so that the traditional information theory-based metrics used in GG02 are not enough to comprehensively estimate the performance of our scheme. Hence, several machine learning-based metrics need to be introduced.

\begin{figure}
\centering
\includegraphics[width=2.2in,height=3in]{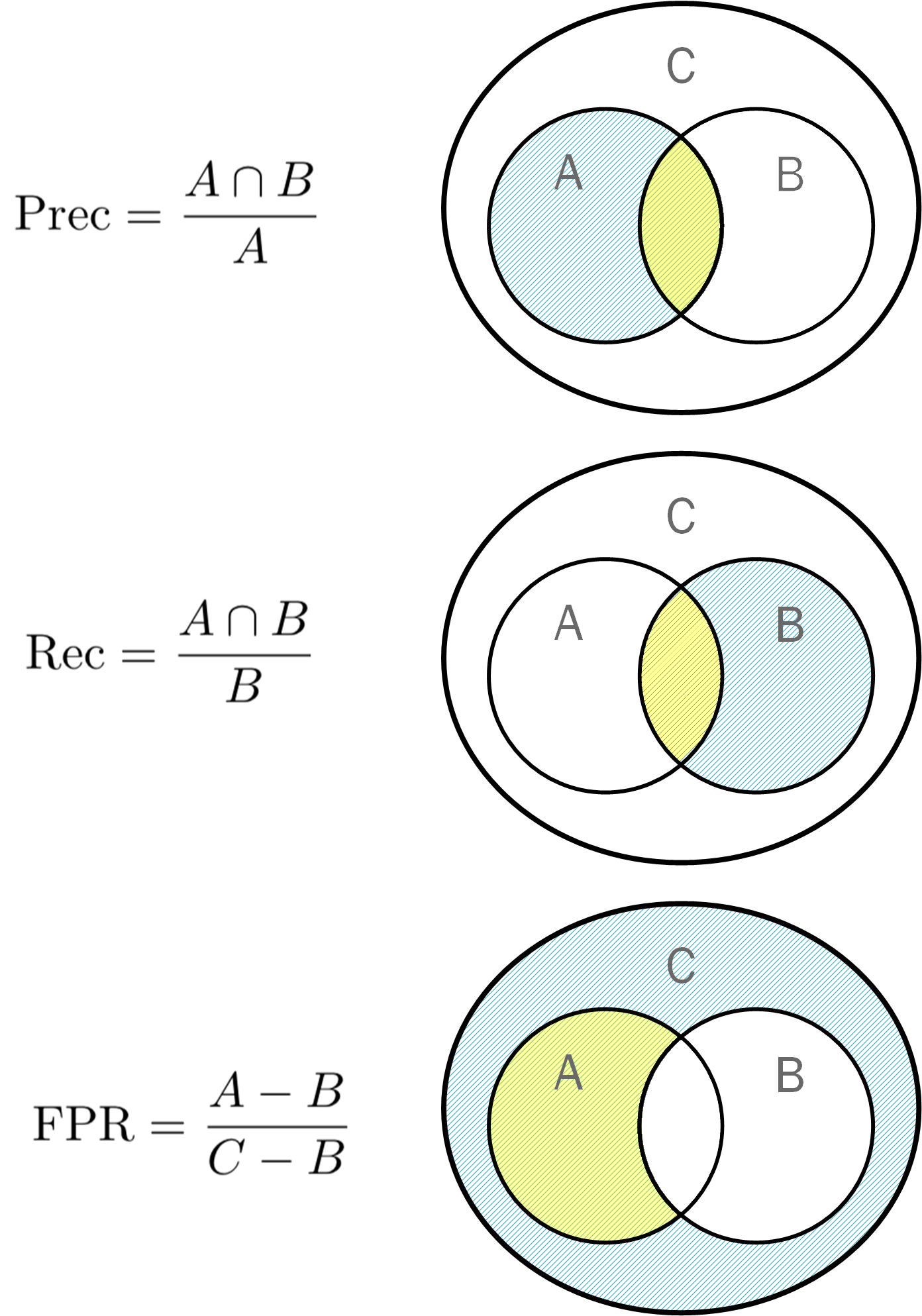}
\caption{The relationship of different datasets and their corresponding metrics.}
\label{fig:sets}
\end{figure}

Assuming there are three datasets, i.e.,  dataset A denotes samples which predicted as positive, dataset B denotes all positive samples, and dataset C denotes all samples. Fig. \ref{fig:sets} shows the relationship between these datasets and their corresponding metrics, which are listed below.

\textit{Precision} (Prec) describes the rate of positive samples in the samples which predicted as positive.

\textit{Recall} (Rec) describes the rate of samples which predicted as positive in all positive samples.

\textit{False Positive Rate} (FPR) describes the rate of samples which predicted as positive in all negative samples.

Actually, besides the above-listed metrics, there may be other metrics used in machine learning to estimate specific system. The reason why we select these three is that the primary concern of our scheme is the correctness of the coherent state classification. We need to know how accurate QMLC can be and how many misclassifications it occurs. In addition, since QMLC is designed for solving multi-label classification problem, we deploy another metric called \textit{Average Precision}  (AP), which evaluates the average fraction of labels ranked above a particular label $y\in Y$ which are in $Y$. It can be expressed as   
\begin{equation}
\begin{aligned}
\mathrm{AP}&=\frac{1}{g}\sum_{i=1}^{g}\frac{1}{|Y_i|} \\
& \times
\sum_{y\in Y_i}\frac{|\{y'|rank_f(|\bm{x_i}\rangle,y')\le rank_f(|\bm{x_i}\rangle,y),y'\in Y_i\}|}{rank_f(\bm{|x_i}\rangle,y)},
\end{aligned}
\end{equation}
where $g$ is the number of data in testing set and $rank_f(\cdot,\cdot)$ is ranking function related to labels \cite{Zhang2007ML}. The performance improves with the increased AP, and the maximum perfect value is $\mathrm{AP}=1$. 

\begin{figure*}
\centering
\includegraphics[width=6.5in,height=4.7in]{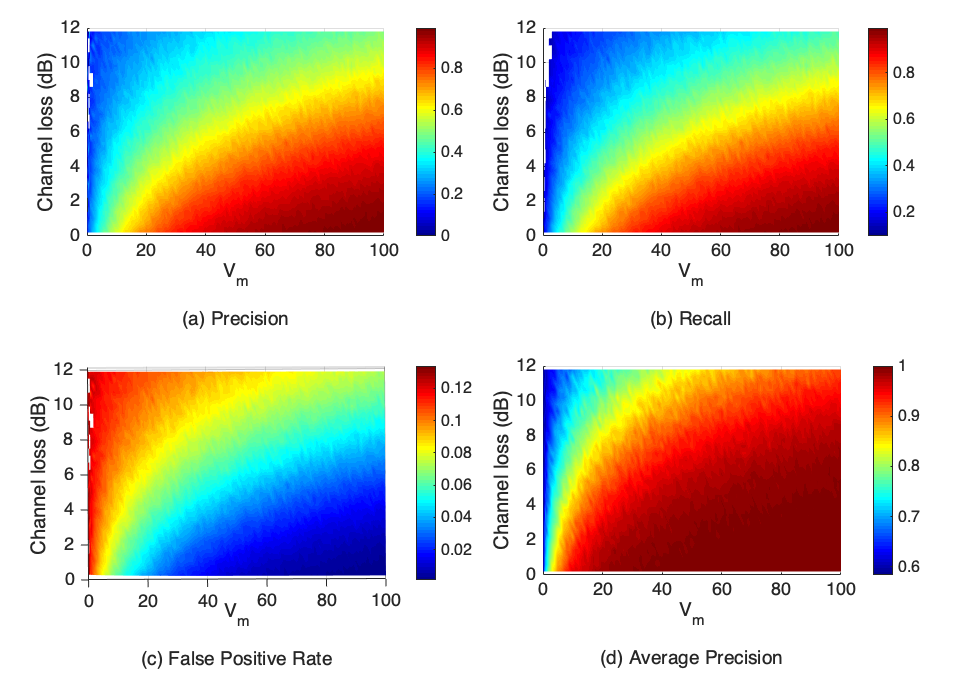}
\caption{Performance of the QMLC-embedded ML-CVQKD system in terms of (a)Precision, (b)Recall, (c)FPR and (d)Average Precision. The number of nearest neighbors is $k=9$ and excess noise is $\xi=0.01$. The training set contains 5000 data points and the testing set contains $10^4$ data points.
}
\label{fig:subplot}
\end{figure*}

Fig. \ref{fig:subplot} shows the performance of the QMLC-embedded ML-CVQKD system in terms of precision (a), recall (b), false positive rate (c) and average precision (d). According to the plots (a), (b) and (d), the performance of Prec/Rec/AP show the similar trend. Namely, it increases with the enlarged modulation variance and decreases with the risen channel loss. More specifically, the optimal  performance can be achieved in both Prec and Rec when $V_m \ge 40$ regulated at a certain channel loss range. It illustrates that the QMLC-embedded ML-CVQKD has the ability to accurately predict unlabeled positive signal states. In the meanwhile, plot (d) shows that the AP has larger range of perfect performance area. It illustrates that QMLC is well qualified for handling the multi-label classification problem of coherent state. On the other hand, plot (c) shows the reduced FPR, and it illustrates that the level of misclassification is well acceptable.

\begin{figure}
\centering
\includegraphics[width=3in,height=2.85in]{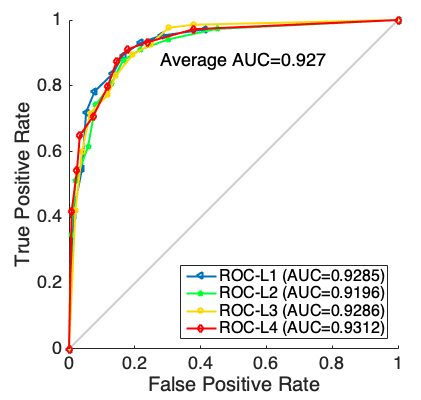}
\caption{ROC curves with respective AUC values of embedded quantum multi-label classifier. Gray line denotes the performance of random guess.}
\label{fig:roc}
\end{figure}

Although Prec/Rec/FPR/AP have shown the respective performance from different aspects, we still hope that using only one metric to check the overall quality of the embedded QMLC. Therefore, \textit{Receiver Operating Characteristic Curve} (ROC) \cite{Cook2007Use}, which describes the true positive rate of a certain classifier as a function of its FPR, is introduced. With ROC curve, one can explicitly tell the quality of the classifier: the curve more close to point (0,1), the performance better. Fig. \ref{fig:roc} shows the ROC curves of QMLC. The gray line is the result of random guess, which illustrates that there is no performance improvement without using any classifier. For each label, however, the ROC curve is close to point (0,1) with embedded QMLC, which denotes the proposed classifier can dramatically improve the prediction performance of ML-CVQKD. We further calculate the \textit{Area Under Curve} (AUC) for each label and thus obtain AUC value, which is a probability value range from 0 to 1. As a numerical value, AUC can be directly used for evaluating classifier's quality. Therefore, the efficiency of quantum classifier $\Lambda$ can be described by its AUC value, namely $\Lambda=$Average AUC in our case. Moreover, a threshold range from 0.5 to 1 can be set to monitor the effectiveness of classifier. The state learning must be interrupted and restarted if AUC value of current trained classifier less than a certain threshold.

\subsection{Security analysis}
Actually, the modulation variance is one of the crucial parameters impacting the performance of CVQKD system. However, the modulation variance $V_m$ of the traditional eight-state protocol cannot be set a larger value, reference \cite{Leverrier:2011cu} points out its optimal value is $V_m=0.7$. Indeed, small modulation variance is usually required for the discretely-modulated CVQKD protocol as it is the only way to prevent information from being eavesdropped by Eve \cite{Leverrier:2009cr}. In the proposed ML-CVQKD, fortunately, this problem can be well solved due to the usage of the multi-label learning framework. As known in the discrete modulation scheme, Eve can intercept modulated coherent states and resend them to Bob without introducing any noise if the modulation variance is large enough. The reason is that the encoding rule of discrete modulation is public for all users, so that Eve can precisely recover the information carried by their intercepted coherent states. In ML-CVQKD, however, only Alice knows the encoding rule at the beginning, and Bob will learn it at the end of state learning. Therefore, even Eve intercepts the modulated coherent states, she cannot exactly recover information without knowing any encoding rule. Even if encoding rule is compromised somehow, Alice and Bob can share a new encoding rule by restarting the state learning. Hence, the security of our scheme can still be guaranteed even if modulation variance is large. An example for interpreting the security of ML-CVQKD is explicitly given in the Appendix \ref{exa}.

Till now, we have demonstrated the performance of QMLC-embedded ML-CVQKD system in terms of machine learning-based metrics and have interpreted its security through an example. However, we still want to present a performance comparison for ML-CVQKD in traditional way so that researchers who do not familiar with machine learning can immediately evaluate how much improvement of the proposed scheme can achieve. To this end, we first present the theoretical security proof for ML-CVQKD as follows.

As known in finite-size scenario, the secret key rate of the traditional CVQKD is given by \cite{Leverrier:2010es}
\begin{equation}\label{ff}
\begin{aligned}
K_{\mathrm{fini}}=\frac{n}{N}[\beta I(A:B)-S_{\epsilon_{PE}}(E:B)-\Delta(n)],
\end{aligned}
\end{equation}
where $\beta$ is the efficiency for reverse reconciliation and $I(A:B)$ is the Shannon mutual information between Alice and Bob and should be computed for a binary-input additive white Gaussian noise (AWGN) channel \cite{Fossier:2009dz}. For heterodyne detection, we have:
\begin{equation}\label{ffi}
\begin{aligned}
I(A:B)=\mathrm{log_2}\frac{V+\chi_{tot}}{1+\chi_{tot}},
\end{aligned}
\end{equation}
where $V=V_m+1$ and the total noise referred to the channel input is $\chi_{tot}=\xi-1+2(1+v_{el})/(\eta T)$, $\eta$ and $v_{el}$ are the practical detector's efficiency and noise due to detector electronics, respectively. The value $N$ denotes the total number of the exchanged signals and $n$ denotes the number of signals that is used for sharing key between Alice and Bob. The remained $N-n$ signals is  used for parameter estimation with failure probability is $\epsilon_{PE}$. The parameter $S_{\epsilon_{PE}}(E:B)$ represents the Holevo bound \cite{Nielsen2011Quantum} of the mutual information between Eve and Bob and needs to be calculated in parameter estimation. The parameter $\Delta(n)$ is related to the security of the privacy amplification, which is given by
\begin{equation}
\begin{aligned}
\Delta(n)=(2\mathrm{dim}\mathcal{H}_{B}+3)\sqrt{\frac{\mathrm{log}_2(2/\bar{\epsilon})}{n}}+\frac{2}{n}\mathrm{log}_2(1/\epsilon_{PA}),
\end{aligned}
\end{equation}
where $\bar{\epsilon}$ is a smoothing parameter, $\epsilon_{PA}$ is the failure probability of privacy amplification, and $\mathcal{H}_{B}$ is the Hilbert space corresponding to the Bob's raw key. Since the raw key is usually encoded on binary bits, we have $\mathrm{dim}\mathcal{H}_{B}=2$. In ML-CVQKD, however, due to the data processing is quite different, Eq.\ref{ff} can be rewritten as the following form
\begin{equation}\label{fff}
\begin{aligned}
K_{\mathrm{fini}}^{\mathrm{ML}}=\frac{n}{N}[\beta\Lambda_{QMLC} I(A:B)-\chi^{ML}_E-\Delta(n)].
\end{aligned}
\end{equation}
The difference between Eq.\ref{ff} and Eq.\ref{fff} lies in two parts. First, the efficiency of embedded quantum classifier $\Lambda$ has to be considered since quantum classifier is necessary to the proposed ML-CVQKD. Note that the QMLC is one of quantum multi-label classifiers suggested in this paper, other excellent quantum classifiers may also fit for ML-CVQKD. Second, term $S_{\epsilon_{PE}}(E:B)$ in Eq.\ref{ff} is substituted by term $\chi^{ML}_E$, which denotes the Holevo quantity of the useful information Eve acquired by interacting with the quantum states during state prediction, reads
\begin{equation}\label{hh}
\begin{aligned}
\chi^{ML}_E=S(\rho_E)-\sum_{y_i}^{m}p(y_i)S(\rho_{E|y_i}).
\end{aligned}
\end{equation}
where $S(\rho)=-\mathrm{Tr}(\rho\mathrm{log}\rho)$ is the von Neumann entropy, the logarithms are taken in base 2, $y_i$ is the raw key obtained by Bob's measurement with probability $p(y_i)$, $\rho_{E|y_i}$ is the corresponding state of Eve's ancilla, and 
\begin{equation}\label{hhh}
\begin{aligned}
\rho_E=\sum_{y_i}^mp(y_i)\rho_{E|y_i}.
\end{aligned}
\end{equation}
In traditional discretely-modulated CVQKD protocol, $p(y_i)=1/m$ is the probability of discrete uniform distribution when variable $Y=y_i$ $(i=1,2,...,m)$, since $Y$ contains $m$ finite and complete encoding events randomly chosen by Alice. For example, $p(y_i)=1/4$ $(m=4)$ in four-state protocol and $p(y_i)=1/8$ $(m=8)$ in eight-state protocol. However, in the case of ML-CVQKD, Eve actually does not know how many encoding events can be chosen by Alice because the encoding rule is changeable and private for her. That is to say, the possible encoding events $Y$ is infinite $(m\rightarrow\infty)$ for Eve, so that an intercepted state could denote any bit(s), which leading $p(y_i)\rightarrow0$. Therefore, Eve can hardly obtain useful information from the intercepted state. The rest calculations can be found in appendix \ref{cal}.

\begin{figure}
\centering
\includegraphics[width=3.4in,height=2.5in]{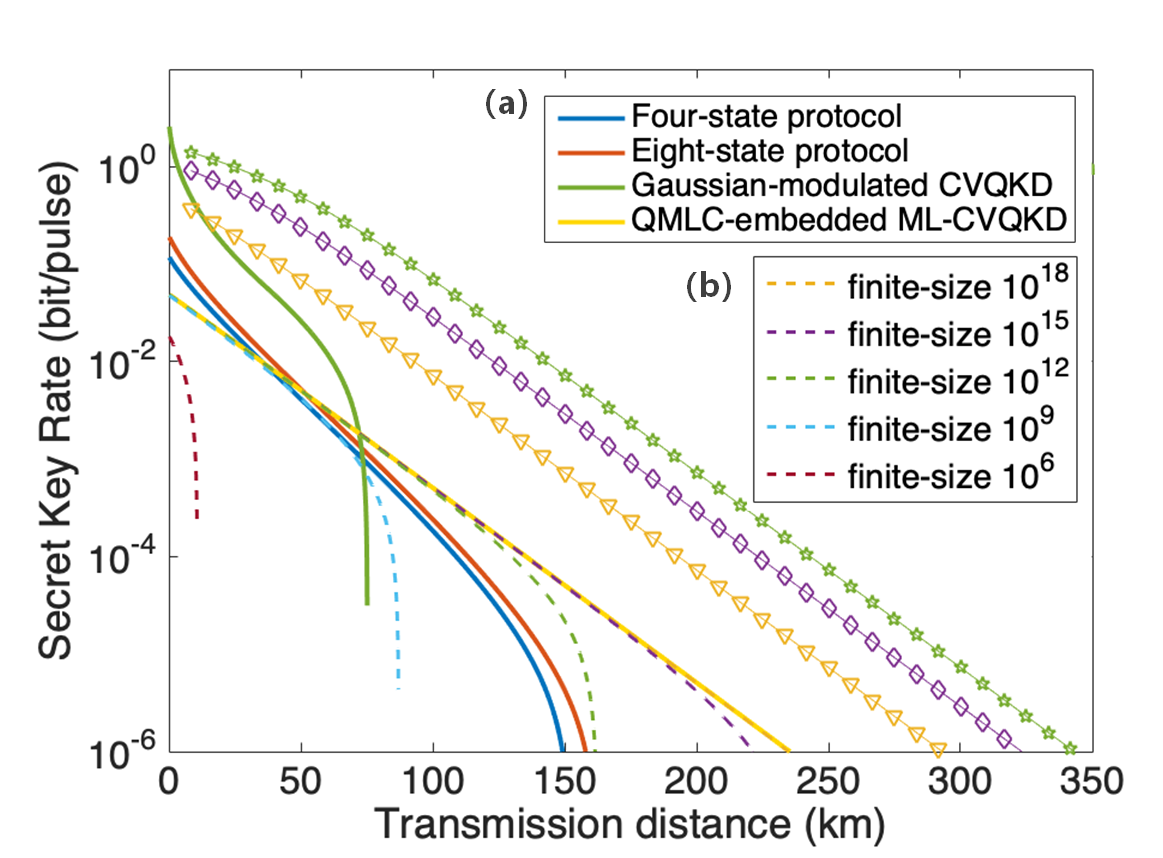}
\caption{Performance comparison. (a) Solid lines denote the asymptotic secret key rates as a function of transmission distance, in which modulation variances $V_m$ are optimized for reasonable SNR \cite{Fossier:2009dz}. (b) Dashed lines denote finite-size secret key rates of QMLC-embedded ML-CVQKD as a function of transmission distance with smallest optimal modulation variance $V_m$=0.35. Lines with triangles, diamonds and pentagrams denote the asymptotic performance of QMLC-embedded ML-CVQKD with $V_m=5$, $V_m=20$ and $V_m=50$, respectively. The parameters are set to $\eta=0.6, v_{el}=0.05$, reconciliation efficiency $\beta=0.98$, classification efficiency $\Lambda=0.927$, excess noise $\xi=0.01$, $n=N/2$ and $\bar{\epsilon}=\epsilon_{PE}=\epsilon_{PA}=10^{-10}$ \cite{Leverrier:2010es}.}
\label{fig:finite1}
\end{figure}
Fig.\ref{fig:finite1}(a) shows the performance comparison between QMLC-embedded ML-CVQKD and several existing CVQKD protocols in asymptotic limit. The result shows that the maximum transmission distance of the proposed scheme outperforms other CVQKD protocols. Fig.\ref{fig:finite1}(b) shows the lower bound of finite-size performance of QMLC-embedded ML-CVQKD where $Vm=0.35$. This value is the smallest optimal modulation variance in eight-state protocol shown in Appendix \ref{cal}. Since small variance is no longer required for the security of ML-CVQKD, the performance of ML-CVQKD can be further increased with the increase of modulation variance. To verify this analysis, we further plot the asymptotic performance of our scheme with larger modulation variances, lines with triangles ($V_m=5$), diamonds ($V_m=20$) and pentagrams ($V_m=50$) show that the maximal secret key rate can also be enhanced with the increase of modulation variance, so that the proposed scheme can outperform traditional discretely-modulated CVQKD protocol in terms of both speed and distance.

\subsection{Practicality}
Finally, let us consider the practicality of ML-CVQKD. At the initial state learning, Alice and Bob are starting to establish a classification model, this process may be a little bit costly since numbers of coherent states and classical data should be communicated and computed. Despite that, the cost is acceptable because only four labels need to be considered, this magnitude is very small for addressing multi-label learning problem \cite{Boosting}. Once the process successfully completed, the system begin to enter the real quantum key distribution process, i.e. state prediction process. This process is more economic than other existing CVQKD protocols, Fig.\ref{fig:finite1}(b) shows that ML-CVQKD still can generate positive secret key when block length is $10^6$, while the secret key rate is null for a block length of $10^6$ in both four-state protocol and eight-state protocol \cite{Leverrier:2010es}. Moreover, the minimum block length of ML-CVQKD for generating positive secret key can also be further decreased with the increase of modulation variance, thereby saving more computational resources. In addition, ML-CVQKD can be applied to the existing optical communication system without any extra equipment, leading to fast deployment and operation.

\section{Conclusion}

In this work, we have proposed a multi-label learning-based scheme for discretely-modulated CVQKD protocol, called ML-CVQKD. In particular, the proposed scheme including two parts which are state learning and state prediction, respectively. State learning is used for training and estimating quantum classifier, while state prediction is used for generating final secret key. To this end, feature extraction was suggested to better represent the characteristics of modulated coherent state. Subsequently, a specialized quantum multi-label classification algorithm (QMLC) was elegantly designed as an embedded classifier for distinguishing the incoming signal state. We then introduced a series of related machine learning-based metrics to estimate the performance of QMLC, and presented the theoretical security proof of ML-CVQKD in both asymptotic limit and finite-size regime. The practicality of ML-CVQKD was also discussed.

Performance analysis shows that QMLC-embedded ML-CVQKD is well feasible and effective for predicting the unknown signal state. We find that ML-CVQKD is able to immune intercept-resend attack, thereby improving the performance of discretely-modulated CVQKD system by take advantages of QMLC classifier. Numerical simulation shows that the proposed QMLC-embedded ML-CVQKD outperforms other existing CVQKD protocols specially in maximum transmission distance, and the performance of both transmission distance and secret key rate will be further increased with the increase of modulation variance.

ML-CVQKD is not only a kind of variant of CVQKD protocol, but also provides a novel thought for introducing various machine learning-based methodologies to CVQKD field.

\begin{acknowledgments}
Q. Liao would like to thank Prof. X. Fu, Prof. X. Wang, Dr. T. Wang and Dr. C. Wang for the helpful discussions. This work is supported by the National Natural Science Foundation of China (Grants Nos. 61572529, 61871407) and the Fundamental Research Funds for the Central Universities.
\end{acknowledgments}

\begin{appendix}

\section{An example for demonstrating the security of ML-CVQKD}\label{exa}

\begin{table*} \label{tab1}
\scriptsize
\centering 
\caption{Encoding rules in different scenarios.}  
\begin{tabular}{lllllllll}  
\toprule 
 & $|\alpha_1\rangle$  & $|\alpha_2\rangle$& $|\alpha_3\rangle$ & $|\alpha_4\rangle$& $|\alpha_5\rangle$& $|\alpha_6\rangle$& $|\alpha_7\rangle$& $|\alpha_8\rangle$\\  
\hline  
Encoding rule 1: eight-state CVQKD (fixed, public) & 000 & 001 & 010 & 011 & 100 & 101 & 110 & 111 \\
Encoding rule 2: ML-CVQKD with state learning 1 (changeable, private)& 111 & 110 & 101 & 100 & 011 & 010 & 001 & 000 \\
Encoding rule 3: ML-CVQKD with state learning 2 (changeable, private)& 00& 10101&11 &1 &1001 &01 &1011 &101 \\

\bottomrule
\end{tabular}  

\scriptsize
\centering
\caption{Decoding results of Alice randomly sends $|\alpha_4\rangle$, $|\alpha_7\rangle$ and $|\alpha_2\rangle$ to Bob.}
\label{Tab03}
\begin{tabular}{cccccccccc}
\toprule
\multirow{2}{*}{ } & \multicolumn{3}{c}{Alice} & \multicolumn{3}{c}{Eve} & \multicolumn{3}{c}{Bob} \\
\cmidrule(r){2-4} \cmidrule(r){5-7} \cmidrule(r){8-10}
&  $|\alpha_4\rangle$      &  $|\alpha_7\rangle$   &   $|\alpha_2\rangle$
&  $|\alpha_4\rangle$      &  $|\alpha_7\rangle$   &   $|\alpha_2\rangle$
&  $|\alpha_4\rangle$      &  $|\alpha_7\rangle$   &   $|\alpha_2\rangle$  \\
\midrule
eight-state CVQKD&011&110&001&011&110&001&011&110&001\\
ML-CVQKD after state learning 1&100&001&110&011&110&001&100&001&110 \\
ML-CVQKD after state learning 2 (encoding rule 1 is compromised)&1&1011&10101&100&001&110&1&1011&10101 \\

\bottomrule
\end{tabular}
\end{table*}

As shown in Tab.I, the encoding rule in discretely-modulated CVQKD is fixed and public, while it can be changed in ML-CVQKD by restart state learning and only known by Alice at first. Tab.II shows the decoding results caused by different encoding rules depicted in Tab.I. Assuming Alice randomly sends $|\alpha_4\rangle$, $|\alpha_7\rangle$ and $|\alpha_2\rangle$ to Bob, and Eve has the ability to totally intercept and resend these signal states without introducing any noise. For eight-state CVQKD, Eve can precisely recover secret key according to the public encoding rule 1 (Alice, Bob and Eve share an identical secret key 001110001). While for the proposed ML-CVQKD, Eve cannot correctly decode secret key from intercepted states since only Alice and Bob know the encoding rule 2 after state learning 1 (Alice and Bob share an identical secret key 100001110, Eve obtains a false secret key 011110001 if she decodes with public encoding rule 1 as before). Even if encoding rule 2 is compromised, a new encoding rule 3 can be generated by restarting state learning 2. As a result, Eve still cannot obtain the correct secret key (Alice and Bob share an identical secret key 1101110101, while Eve obtains a false secret key 100001110 if she decodes with compromised encoding rule 2). Moreover, as shown in encoding rule 3, the encoding length of ML-CVQKD is variable and, theoretically, it even can be set to arbitrary length. Therefore, Eve becomes more difficult to obtain correct secret key from its intercepted states in ML-CVQKD. 

\section{Calculation for discretely-modulated CVQKD protocol}\label{cal}
Here, we present the calculation of asymptotic secret key rate of discretely-modulated CVQKD protocol. Its finite-size case can be found in \cite{GuoPerformance}.

The the asymptotic secret key rate of CVQKD protocol where Bob performs heterodyne detection with reverse reconciliation under collective attack can be given by
\begin{equation}\label{wew}
\begin{aligned}
K_{asym}=\beta I(A:B)-\chi_{BE},
\end{aligned}
\end{equation}
where the mutual information of Alice and Bob $I(A:B)$ is already given by Eq. (\ref{ffi}), and
\begin{equation}\label{wewW}
\begin{aligned}
\chi_{BE}=\sum_{i=1}^2G\left(\frac{\lambda_i-1}{2}\right)-\sum_{i=3}^5G\left(\frac{\lambda_i-1}{2}\right),
\end{aligned}
\end{equation}
where $G(x)=(x+1)\mathrm{log_2}(x+1)-x\mathrm{log_2}x$ is the von Neumann entropy, and the symplectic eigenvalues
\begin{equation}\label{app1}
\begin{aligned}
\lambda_{1,2}^2=\frac{1}{2}[A\pm\sqrt{A^2-4B}]  
\end{aligned}
\end{equation}
with
\begin{equation}
\begin{aligned}
A=V^2+T^2(V+\chi_{line})^2-2TZ^2
\end{aligned}
\end{equation}
and
\begin{equation}
\begin{aligned}
B=T^2(V^2+V\chi_{line}-Z^2)^2,
\end{aligned}
\end{equation}
\begin{equation}\label{app2}
\begin{aligned}
\lambda_{3,4}^2&=\frac{1}{2}[C\pm\sqrt{C^2-4D}]
\end{aligned}
\end{equation}
with
\begin{equation}
\begin{aligned}
C&=\frac{1}{T^2(V+\chi_{tot})^2}[A\chi_{het}^2+B+1 \\
&+2\chi_{het}(V\sqrt{B}+T(V+\chi_{line})+2TZ^2)] 
\end{aligned}
\end{equation}
and
\begin{equation}
\begin{aligned}
D&=\left(\frac{V+\sqrt{B}\chi_{het}}{T(V+\chi_{tot})}\right)^2,
\end{aligned}
\end{equation}
$\lambda_5=1$, where the total channel-added noise referred to the channel input is defined as $\chi_{line}=1/T-1+\xi$ and the detection-added noise referred to Bob's input for heterodyne detection is $\chi_{het}=[1+(1-\eta)+2v_{el}]/\eta$. The term $Z$ is the correlation between Alice and Bob. In Gaussian-modulated CVQKD protocol, $Z$ usually equals to $Z_G=\sqrt{V^2-1}$. However, In discretely-modulated CVQKD, $Z$ is quite different. Specifically, $Z_4$ is the correlation between Alice and Bob in four-state CVQKD and $Z_8$ is the correlation between Alice and Bob in eight-state CVQKD, which can be expressed by \cite{Leverrier:2011cu}
\begin{equation}\label{z4}
\begin{aligned}
Z_4&=2\alpha^2\sum_{k=0}^3\frac{l_{k-1}^{3/2}}{l_k^{1/2}} 
\end{aligned}
\end{equation}
with
\begin{equation}
\begin{aligned}
l_{0,2}&=\frac{1}{2}e^{-\alpha^2}[\mathrm{cosh}(\alpha^2)\pm\mathrm{cos}(\alpha^2)],
\end{aligned}
\end{equation}
\begin{equation}
\begin{aligned}
l_{1,3}&=\frac{1}{2}e^{-\alpha^2}[\mathrm{sinh}(\alpha^2)\pm\mathrm{sin}(\alpha^2)].
\end{aligned}
\end{equation}
And
\begin{equation}
\begin{aligned}
Z_8&=2\alpha^2\sum_{k=0}^7\frac{l_{k-1}^{3/2}}{l_k^{1/2}} 
\end{aligned}
\end{equation}
with
\begin{equation}
\begin{aligned}
l_{0,4}&=\frac{1}{4}e^{-\alpha^2}[\mathrm{cosh}(\alpha^2)+\mathrm{cos}(\alpha^2)\pm2\mathrm{cos}\left(\frac{\alpha^2}{\sqrt{2}}\right)\mathrm{cosh}\left(\frac{\alpha^2}{\sqrt{2}}\right)],
\end{aligned}
\end{equation}
\begin{equation}
\begin{aligned}
l_{1,5}&=\frac{1}{4}e^{-\alpha^2}[\mathrm{sinh}(\alpha^2)+\mathrm{sin}(\alpha^2) \pm\sqrt{2}\mathrm{cos}\left(\frac{\alpha^2}{\sqrt{2}}\right)\mathrm{sinh}\left(\frac{\alpha^2}{\sqrt{2}}\right)] \\
 &\pm\sqrt{2}\mathrm{sin}\left(\frac{\alpha^2}{\sqrt{2}}\right)\mathrm{cosh}\left(\frac{\alpha^2}{\sqrt{2}}\right)],
\end{aligned}
\end{equation}
\begin{equation}
\begin{aligned}
l_{2,6}&=\frac{1}{4}e^{-\alpha^2}[\mathrm{cosh}(\alpha^2)-\mathrm{cos}(\alpha^2)\pm2\mathrm{sin}\left(\frac{\alpha^2}{\sqrt{2}}\right)\mathrm{sinh}\left(\frac{\alpha^2}{\sqrt{2}}\right)],
\end{aligned}
\end{equation}
\begin{equation}
\begin{aligned}
l_{3,7}&=\frac{1}{4}e^{-\alpha^2}[\mathrm{sinh}(\alpha^2)-\mathrm{sin}(\alpha^2) \mp\sqrt{2}\mathrm{cos}\left(\frac{\alpha^2}{\sqrt{2}}\right)\mathrm{sinh}\left(\frac{\alpha^2}{\sqrt{2}}\right)] \\
 &\pm\sqrt{2}\mathrm{sin}\left(\frac{\alpha^2}{\sqrt{2}}\right)\mathrm{cosh}\left(\frac{\alpha^2}{\sqrt{2}}\right)]. 
\end{aligned}
\end{equation}

\begin{figure}
\centering
\includegraphics[width=3.3in,height=2.4in]{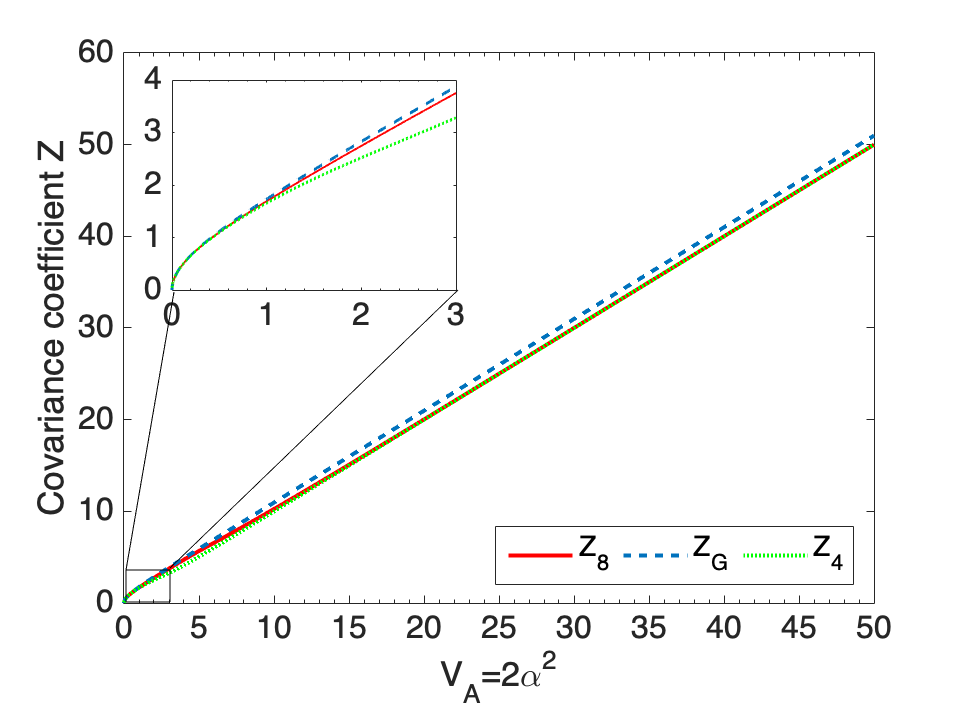}
\caption{Comparison of the covariance coefficient $Z$ as a function of the variance of modulation $V_A$, where $Z_G$ for Gaussian-modulated CVQKD protocol, $Z_4$ for four-state CVQKD protocol and $Z_8$ for eight-state CVQKD protocol.}
\label{zzzz}
\end{figure}
Fig.\ref{zzzz} shows the comparison of different covariance coefficients as a function of modulation variance $V_A$. We find that $Z_4$ and $Z_8$ are equal to $Z_G$ when the modulation variance is small enough. Hence, for a sufficiently low modulation variance the bound $\chi_{BE}$ for discrete modulation is almost identical to the one obtained for a Gaussian modulation.

\begin{figure}
\centering
\includegraphics[width=3.3in,height=2.4in]{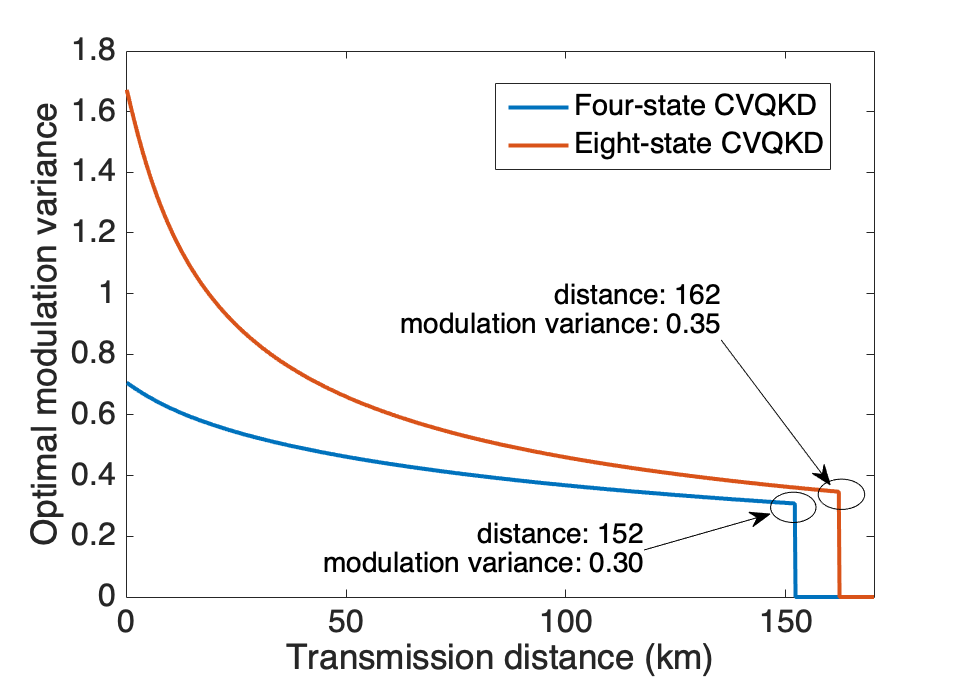}
\caption{Optimal modulation variance for discretely-modulated CVQKD protocol as a function of transmission distance. The parameters are set to $\eta=0.6$, $v_{el}=0.05$, reconciliation efficiency $\beta=0.98$ and excess noise $\xi=0.01$.}
\label{fe}
\end{figure}
Fig.\ref{fe} depicts the optimal modulation variance for discretely-modulated CVQKD protocol as a function of transmission distance. As can be seen from the figure, the optical modulation variance is decreased with the increase of transmission distance, and the minimum optimal modulation variances are 0.3 for four-state CVQKD and 0.35 for eight-state CVQKD. This numerical simulation shows that small modulation variance is required to guarantee safety for discretely-modulated CVQKD protocol as it can prevent eavesdropper from intercepting the useful information.

\end{appendix}

\bibliography{mybibfile.bib}

\end{document}